# Intelligent Perception Assisted Light Modulation for real-time and program-free nanofabrication and particle manipulation

Weixin Zhu [1,#], Hanyu Guo [1,#], Chen Zhang [1], Fang Yang [2], Ce Zhang [1], Lijing Zhong [3,*], Jianrong Qiu [4], Kaige Wang [1,*], Jintao Bai [1], Wei Zhao [1,*]

[1] State Key Laboratory of Photon-Technology in Western China Energy, International Collaborative Center on Photoelectric Technology and Nano Functional Materials, Laboratory of Optoelectronic Technology of Shaanxi Province, Institute of Photonics & Photon Technology, Northwest University, Xi'an 710069, China

[2] Key Laboratory for Molecular Enzymology and Engineering of Ministry of Education, School of Life Sciences, Jilin University, Changchun 130012, China

[3] Institute of Light+X Science and Technology, Faculty of Electrical Engineering and Computer Science, Ningbo University, Ningbo 315211, China

[4] State Key Laboratory of Extreme Photonics and Instrumentation, College of Optical Science and Engineering, Zhejiang University, Hangzhou 310027, China

Correspondence: zhonglijing@nbu.edu.cn; wangkg@nwu.edu.cn; zwbayern@nwu.edu.cn

## Abstract

In this paper, we present an **Intelligent Perception-Assisted Light Modulation** (IPALM) technique, for femtosecond laser nanofabrication and particle manipulation. IPALM technique integrates real-time hand-motion recognition with dynamic spatial light modulation to achieve programming-free laser beam control. In contrast to the conventional programmed laser fabrication techniques which need setup a project prior to fabrication, IPALM offers a direct "mind-to-matter" pathway for laser nanostructuring and biological cells handling with high flexibility and multiple degree of freedom. In nanofabrication, IPALM provides high resolution with feature dimensions down to 280 nm. By mind-driven hand gesture control, the laser beam is regulated to enable direct writing of micro/nanostructures with a minimum feature size down to 280 nm via IPALM. In precise particle manipulation, multiple cells can be simultaneously moved to achieve cell coalescence. With these examples, IPALM showcases its potential for applications in photonics, biomedicine, and microfluidics, for high-dimension and flexible laser applications.

## 1. Introduction

Light field modulation technology, as one of the core research directions in modern optics, enables precise control of light beams through phase modulation[1], and polarization manipulation[2], providing powerful tools for applications in micro/nnofabrication[3] and optical micromanipulation[4-6]. Over the past decades, significant advancements have been made in spatial light modulations (SLMs)[7], allowing dynamic control of wavefronts[8] and generation of complex optical fields such as vortex beams and Bessel beams[9,10]. These technologies have revolutionized fields including photonic device manufacturing[11-13], biomedical imaging[14], and optical trapping[15-17].

Despite these advancements, conventional light field modulation approaches face fundamental limitations in real-time adaptability and multi-degree-of-freedom control[18,19]. Traditional SLM-based systems typically require preset Computer-Generated Hologram (CGH) [20,21], which lack flexibility for on-the-fly adjustments during micro/nano fabrication or particle manipulation processes[22-24] . This "rigid" nature often results in inefficient fabrication workflows and limited capabilities in dynamic biological environments[25]. Moreover, existing beam steering techniques using galvanometer scanner or micromirror arrays offer limited parallelization capabilities and cannot generate arbitrary complex field distributions simultaneously[26,27].

The emerging demand for real-time, interactive optical systems in precision manufacturing and biological research has driven the need for innovative solutions that bridge human intuition with optical control[28-30]. Recent developments in machine learning and computer vision have enabled new paradigms for human-machine interfaces[31]. Their integration with optical modulation systems can translate human actions directly into optical responses without extensive programming requirements[32] (Fig. 1a).

## 2. Results

### 2.1 Laser direct writing of nanostructures

The feasibility and performance of the IPALM technique were systematically evaluated through a series of micro/nanofabrication experiments.

Fig. 2a shows the fabrication results with a single finger. A tight laser focus (Fig. 2b1 and Supplementary Video 1) due to TPA can be monitored by the fluorescence of DETA. Accompanied by the motion of the finger, the laser focus moves simultaneously (Supplementary Video 2). Since the high NA feature of 100× oil-immersion objective, the TPA region in a single exposure is quite small. To generate a continuous and smooth structure via the IPALM technique, the movement of finger should be slowly and stably controlled. This is a challenge for human's hand. Therefore, the characters writing by finger is not explicitly controlled as shown in Fig. 2a. However, if we want to draw a smooth line (Fig. 2b), it is still realizable by IPALM technique. The minimum line widths could be as low as 280 nm under 8 mW laser power, indicating the high precision of nanofabrication by IPALM technique.

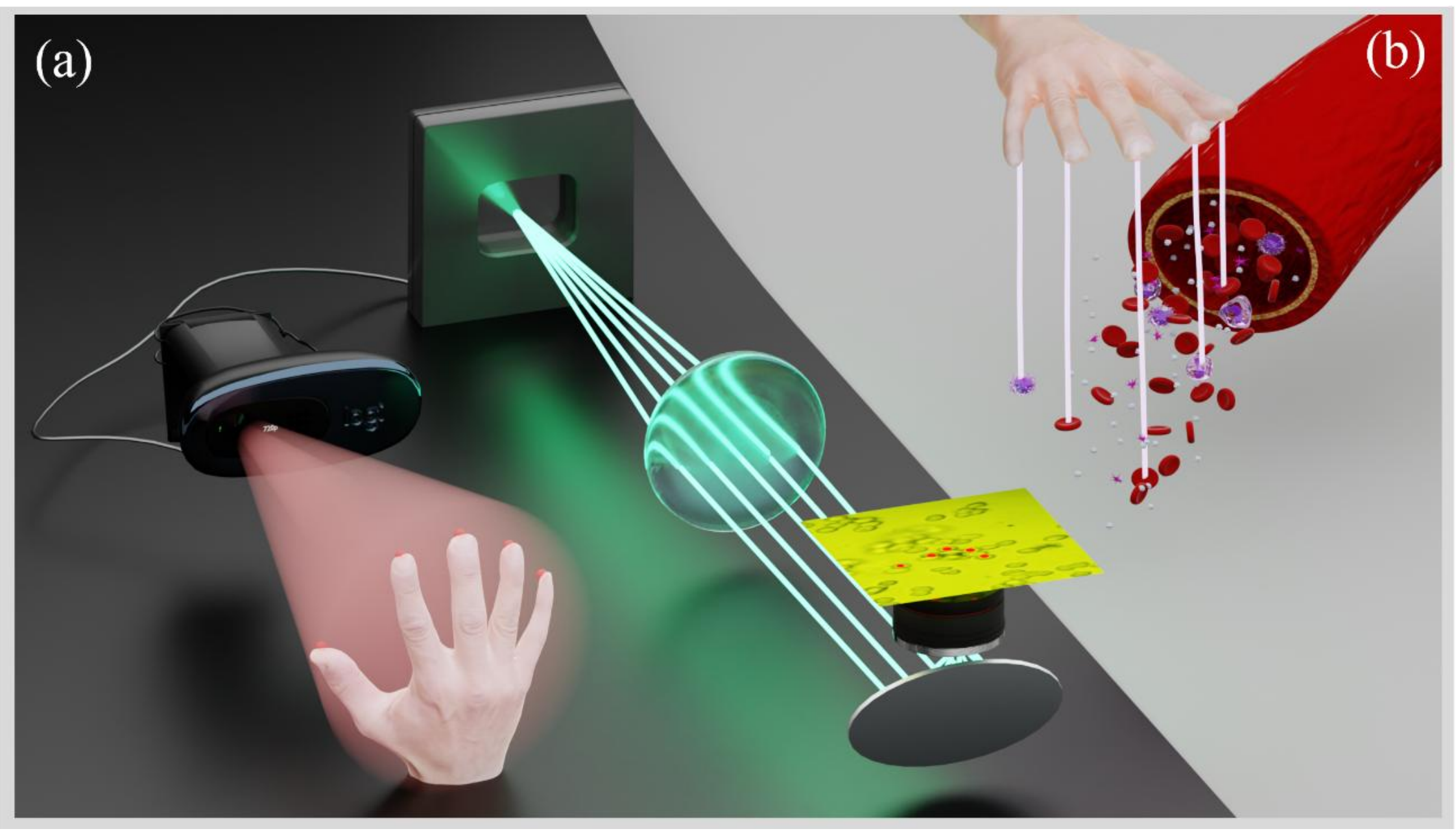

**Fig. 1 Dynamic light field control that integrates real-time hand motion recognition with non-iterative phase pattern generation based on IPALM technology.** (a) System schematic of the IPALM setup, showing the integration of hand tracking and light field control modules. (b) Demonstration of the system's practical application in simulating vascular-scale biological manipulation via intuitive gesture interaction.

We further demonstrated the capability of three- (Fig. 2c, Supplementary Video 3) and five-finger (Fig. 2d, Supplementary Video 4) fabrications of micro/nanostructures through IPALM. Smooth line structures have been parallelly written by the hand motions. For instance, under 15 mW laser power, a minimum line width of 280 nm was achieved (Fig. 2d2).

Further study was conducted to quantify the effects of exposure time and laser power online width. Fig. 2e shows that longer exposure times lead to wider line widths due to increased energy deposition. In the investigated range of parameters, the line width exhibits approximately linear relationship with exposure time and laser power. These are consistent with conventional laser direct writing (LDW), indicating IPALM technique can be employed as conventional LDW.

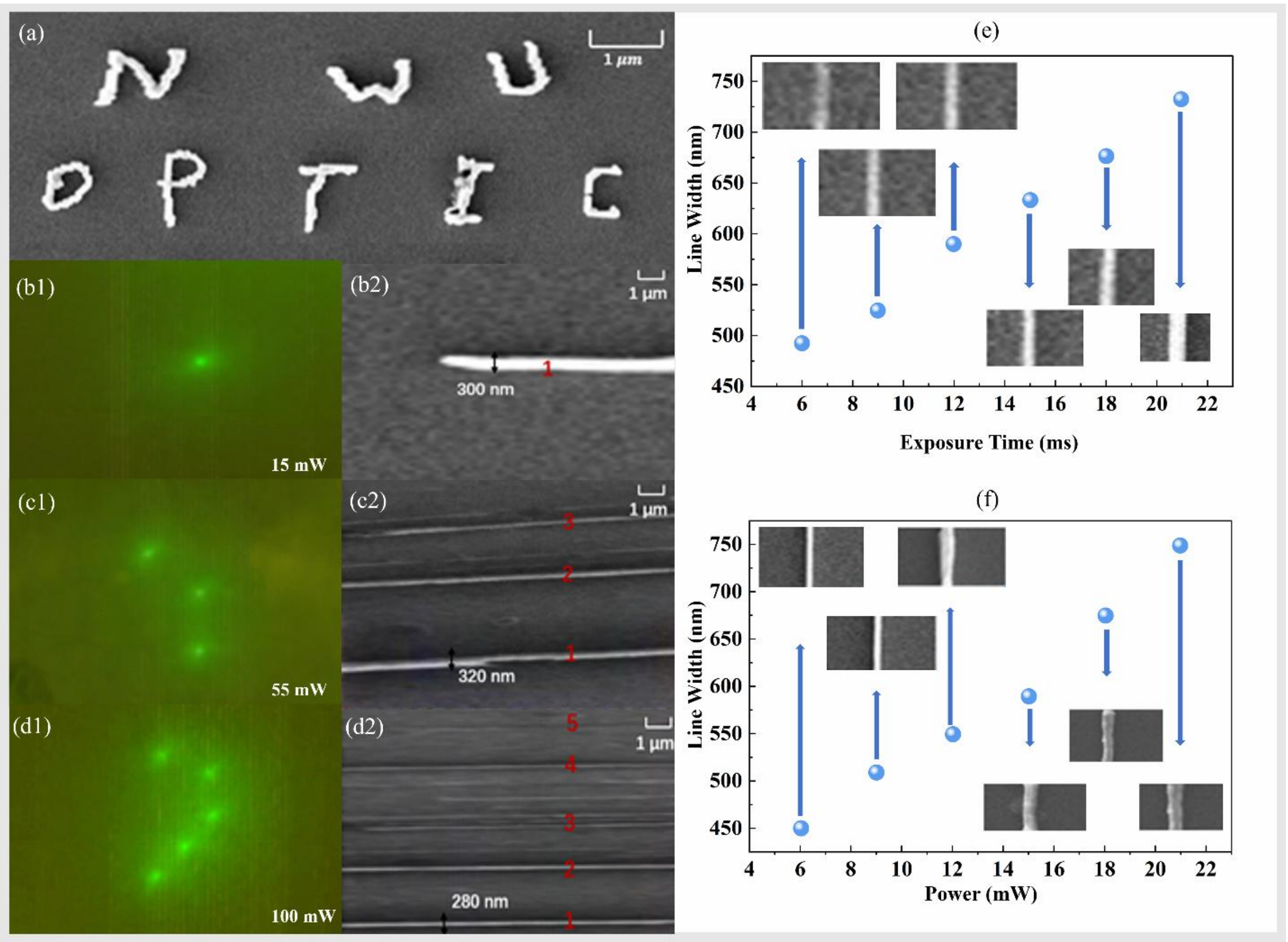

**Fig. 2 Laser direct writing of nanostructures via IPALM.** (a) SEM results of fabricated characters written by IPALM at a power of 15 mW. From left to right, they respectively show the optical fluorescence response observed by the CMOS camera and the SEM scan results of different numbers of modulated light spots under a 100X objective lens. (b1-b2) Modulating a single laser beam at a power of 15 mW and the corresponding SEM result of laser direct writing. (c1-c2) Modulating three laser beams at a power of 55 mW and the corresponding SEM result of laser direct writing. (d1-d2) Modulating three laser beams at a power of 100 mW and the corresponding SEM result of laser direct writing. (e) The influence of different exposure times on the line width of the structure; (f) The influence of different entrance pupil powers on the line width of the structure.

## 2.2 Particle manipulation

*Manipulation of Sheep Red Blood Cells*

In Fig. 3a-d and Supplementary Video 5, we demonstrate sheep red blood cells (RBCs, ~5 μm diameter) were successfully trapped and manipulated using a single Bessel beam generated guided by a single finger. The minimum power threshold for stable trapping and manipulation was found to be ~5.5 mW.

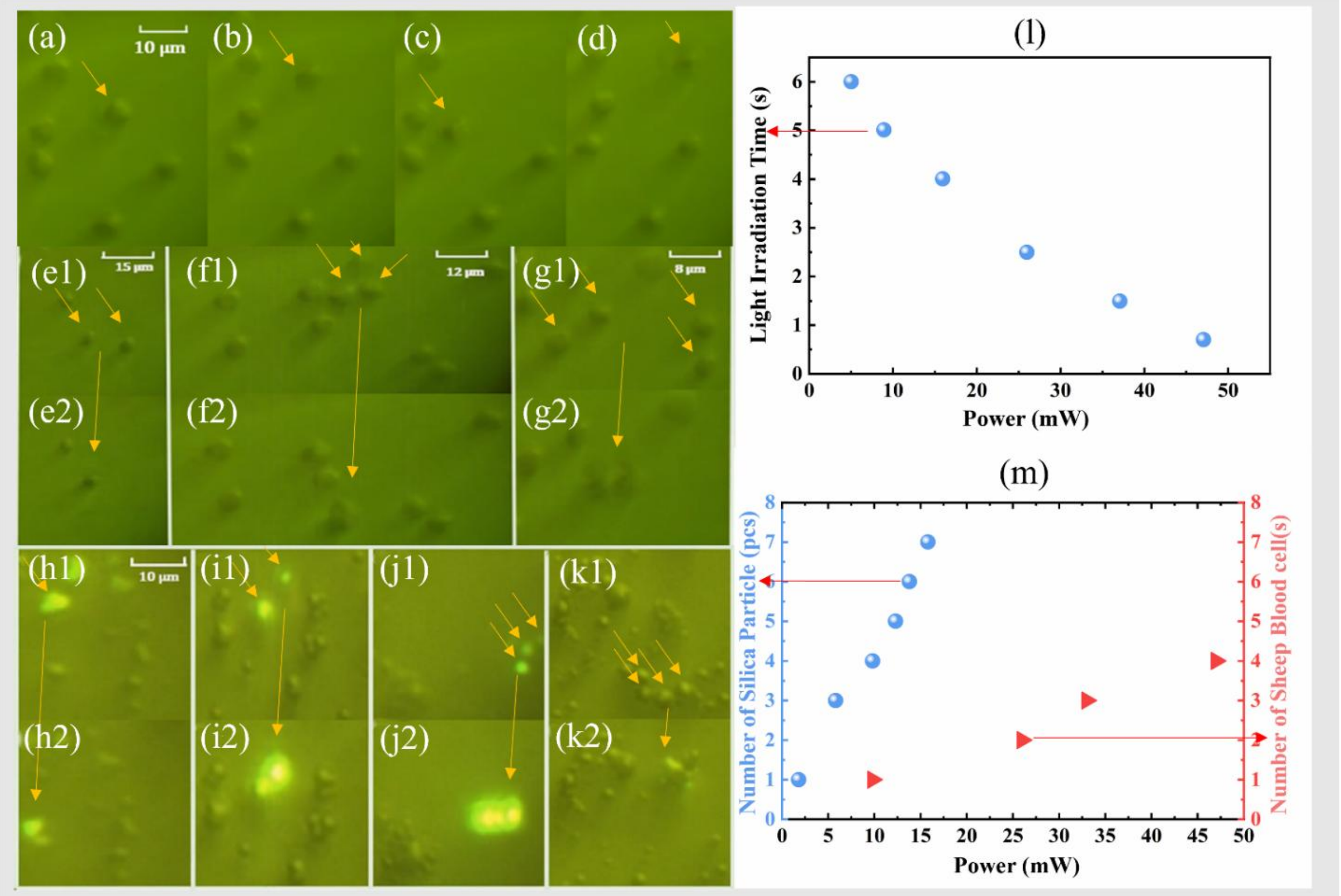

**Fig. 3 Optical manipulation via IPALM.** (a-d) Bright field images of sheep blood cell manipulation by IPALM at a power of 5.5 mW. One single cell highlighted by circle is moved by a single laser beam. (e-g) Multicellular sequential capture and parallel manipulation, with the fixed optical trap position, by increasing the laser power in a stepwise manner, the sequential pulling and stable trapping of multiple sheep red blood cells were achieved. (e) When the laser power is 26 mW, the first free cell is pulled toward the second cell by the optical trap, and eventually both cells are simultaneously trapped by the same optical trap. (f) When the power is increased to 33 mW, the first two trapped cells are pulled toward the third cell as a whole, achieving the synchronous trapping of three cells. (g) When the power is further increased to 47 mW, the first three cells are pulled toward the fourth cell, and finally the stable co-trapping of four cells is completed. (h-k) Sequential capture of multiple particles and array construction. The process of step-by-step capturing of silica microspheres (diameter ~1 μm) using a single moving optical trap is demonstrated, with the same scale bar applied to all subfigures. (h) A single microsphere is successfully trapped at an incident pupil power of 1.8 mW. (i) When the power is increased to 4.2 mW, the optical trap drags the first microsphere to move and successfully captures the second microsphere, forming a dimer. (j) At a power of 5.8 mW, the optical trap drags the first two microspheres and successfully captures the third one, forming a trimer. (k) When the power reaches 9.8 mW, the optical trap drags the first three microspheres and finally captures the fourth one, completing the construction of the tetramer array. (l) Relationship between critical irradiation time for membrane damage and laser power for single beam. (m) Quantitative relationship between capturing capability and laser power.

When the laser power is further increased, multiple RBCs can be simultaneously manipulated via IPALM, as shown in Fig. 3e~g and Supplementary Video 6, where 2, 3 and 4 cells were captured and moved according to the motion of hand. The required laser powers are 26 mW, 33 mW, and 47 mW, respectively.

In the manipulation of RBCs, the selection of laser power is tricky. A high laser power can promote the controllability of cells, while may cause cell damage as well. For instance, when a single Bessel beam is applied at a higher power (>26 mW), cell membrane is damaged under the irradiation. Fig. 3l and Supplementary Video 7 illustrate the relationship between laser power and exposure time leading to membrane rupture. The data clearly show that a longer period of manipulation is only achieved at a lower laser power. Fortunately, the requirement of the threshold of laser power to manipulate multiple cells does not increase drastically with the number of simultaneously trapped cells (Fig. 3m). A nearly linear relationship implies multiple manipulation can be effective and efficiently reached for cell sorting, cancer invasion test and medicine test.

*Manipulation of Silica Microparticles*

Silica particles (2 μm diameter) were also manipulated to demonstrate the general applicability of IPALM. Although the silica particles have larger density than cells, due to its small size and relatively stronger gradient force, the threshold for trapping a single silica particle was **1.8 mW**. In contrast to RBCs, silica particles are difficult to damage under the irradiation of femtosecond laser. They can afford higher laser power, thus, stronger manipulation by IPALM technique. Fig. 3h~k and Supplementary Video 8 show the sequential capture of 1 to 4 particles with increasing laser power. For instance, three silica particles are simultaneously dragged towards the fourth one, forming a tetramer, only at 9.8 mW (Fig. 3j). The laser power required for multiple silica particles manipulation is much lower than RBCs (Fig. 3l).

## 3. Discussions

To this end, we have demonstrated the significant potential of IPALM technology across multiple disciplines, e.g. micro/nanostructure fabrication, cell and particle manipulation, which are applicable in photonics industry, biomedicine, pharmaceutical synthesis, and medicine test.

However, the current technique still faces several challenges. On one hand, for instance, although hand provides a high flexibility and degree of freedom, it cannot provide sufficiently high stability relative to the counterpart realized by machines. This is not a big deal in cell manipulation and the corresponding applications in pharmaceutical synthesis and medicine-cell interaction investigation but do affect the smoothness of fabricated nanostructures. On the other hand, the real-time capability of IPALM technique can still be promoted. Currently, only a laptop computer with low computation capacity has been applied to develop the system. There is also no additional AI technique used to optimize the system. It is not beyond expectation that, with the aid of high-performance computer system and AI technique, the prototype of IPALM technique in this manuscript will have a significant promotion by orders.

## 4. Conclusions

In this study, we have developed an Intelligent Perception-Assisted Light Modulation technique, integrating hand motion recognition, spatial light modulator and non-iterative CGH design algorithms, to reach a real-time, program-free, human-computer interactable light control method. The overall time delay from receiving the live video from the camera to uploading CGH to SLM is only 42 ms, equivalent to 23.8 fps under GPU acceleration. The technique breaks traditional pre-programming limits, enabling "mind-to-matter" interaction.

In nanofabrication, with a 100× oil-immersion objective, IPALM achieves a minimum feature size of 280 nm, suppresses hand tremor noise via filtering. For cell manipulation, it traps sheep RBCs at a 5.5 mW threshold, manipulates 2–4 cells or silica particles simultaneously. IPALM technique has demonstrated potential in micro/nano manufacturing, biomedicine and materials science. It not only provides a new paradigm for light field manipulation but also pioneers the emerging interdisciplinary field of Human-Computer Interactive Optics, potentially serving as a core technological platform for future intelligent laboratories.

## 5. Methods

### 5.1 Intelligent Perception-Assisted Light Modulation System

This study developed an Intelligent Perception-Assisted Light Modulation (IPALM) system that integrates real-time hand motion perception with spatial light modulation to achieve dynamic multi-focus light field generation and control.

The optical system is illustrated in Fig. 4a. A 780 nm femtosecond laser (Coherent Chameleon Ultra II) with 140 fs pulse width and a maximum output power of 3.5 W is used as light source. A power control module that is combined by a Glan prism (Thorlabs AFS-SF10) and half-wave plate (Thorlabs AHWP05M) has been applied for femtosecond laser power adjustment. The laser beam is further expanded by a 4f system (lens LA4130-B), filtered by a spatial light pinhole filter (Thorlabs P2500UK). Then, the laser beam is incident on a phase-only liquid crystal SLM (Holoeye PLUTO-2-VIS-016, 60 Hz refresh rate) which has a water-cooling module. The modulated beam is reflected from the SLM, filtered by a variable aperture (Thorlabs FRSTA-10IB-M) to remove the zeroth-order light. Only the modulated beam is reserved. It is then reflected on a dichroic mirror (Chroma ZT532rdc_NIR), and then delivered into objectives (Olympus, 100X NA 1.25). Both the fabrication and cell manipulation processes are monitored in real-time by a CMOS camera (MER-2000-19U3C-L).

### 5.2 Motion Perception and Recognition

Real-time hand motion capture was achieved through convolutional neural network (CNN) and multi-modal sensor fusion. A 4-megapixel CCD camera has been applied to capture the motion of a hand. The live video is transported to a laptop computer (Lenovo Legion 9000K, Table 1) for motion perception and recognition (Fig. 4b). Since the camera has inevitable image aberration, the live video was refined through checkerboard calibration (Fig. S3b). For fast processing with low requirement on computer system, a lightweight neural network—Caffemodel architecture[33](Fig. S3a) has been applied to recognize the 21 hand key points from each frame of the live video and output their coordinates with pixel-level recognition accuracy.

Table 1. Key hardware and software specifications of the Lenovo Legion R9000K

|  |  |
| --- | --- |
| Processor (CPU) | AMD Ryzen 7 5800H (8 Cores, 16 Threads, 3.2 GHz base, up to 4.4 GHz) |
| Graphics (GPU) | NVIDIA GeForce RTX 3070 Laptop GPU (8 GB GDDR6 VRAM) |
| Camera | 720p high-definition camera (4 megapixels) |

To further reduce the recognition time of hand key points, both the CPU (AMD Ryzen 7 5800H) and GPU (NVIDIA RTX 3070) are collaborated to accelerate the processing. In this investigation, only the five hand key points corresponding to the fingertips are used. The cost of recognition time is approximately 40 ms.

The recognition of these key points always leads to fluctuations of recognition positions[34] (Fig. 4c and Supplementary Video 9), since the hand key points are not ideal points. To overcome the jitter, a four-term exponential moving average median filtering (Equations 2.1–2.5) has been applied, with smoothing factor $\alpha = 0.5$ and window size $k = 5$ to eliminate jitter noise.

$$S_{1_t} = \alpha x_t + (1 - \alpha) S_{1_{t-1}} \tag{2.1}$$

$$S_{2_t} = \alpha S_{1_t} + (1 - \alpha) S_{2_{t-1}} \tag{2.2}$$

$$S_{3_t} = \alpha S_{2_t} + (1 - \alpha) S_{3_{t-1}} \tag{2.3}$$

$$S_{4_t} = \alpha S_{3_t} + (1 - \alpha) S_{4_{t-1}} \tag{2.4}$$

$$y_t = \mathrm{Median}(S_{4_{t-k}}, \dots, S_{4_t}, \dots, S_{4_{t+k}}) \tag{2.5}$$

where $S_{i_t}$ represents the exponential moving average value of the i-th layer and $S_{i_{t-1}}$ represents the smoothed value of the i-th layer at the previous moment. The original initial value $x_t$, and $x_0$ is the first value of the original data sequence.$y_t$ represents a smoothed result that both retains the main trend of the data and eliminates short-term fluctuations. Conversion from image coordinates $(x_{img}, y_{img})$ (Fig. 4c) to objective focal plane relative coordinates $(x_{ob}, y_{ob})$ is used

$$x_{ob} = \frac{x_{img}}{W/2}, \quad y_{ob} = \frac{y_{img}}{H/2} \tag{2.6}$$

where $W$ and $H$ represent the width and height of real-time video images.

**5.3 CGH generation**

In this investigation, the incident laser beam is modulated into Bessel beams according to the number of fingers. A non-iterative Strip Segmentation Phase (SSP) algorithm[35] has been applied to enable real-time generation of CGHs (Fig. 4e). In the design of CGH, the real-time position of each of the five fingertips can be directly reflected by the phase function $\psi_i(x, y)$

$$\psi_i(x, y) = \psi_B + \frac{2\pi}{\lambda}\frac{NA}{Rn_t}(x\Delta x_i + y\Delta y_i) \tag{2.7}$$

where $\psi_B = 2\pi\lambda^{-1}\sqrt{x^2 + y^2}\tan\alpha$ is Bessel phase function, $\alpha$ is the angle of axicon lens, $(x, y)$ represent objective rear aperture coordinates, and $(\Delta x_i, \Delta y_i)$ denote the offset of its beam. $(\Delta x_i, \Delta y_i)$ are linearly correlated to $(x_{ob}, y_{ob})$ and $(x_{img}, y_{img})$ accordingly. Thus, the image plane is linearly projected on the focal plane of objective. A blazed grating has also been superimposed, to move the modulation beams away from the zero-order light. The lateral offset is 0.95 μm, while the vertical offset is 5 μm. With the aid of GPU acceleration, each CGH can be parallelly computed in approximately 1 ms.

Overall, the time expanse from capturing hand motion, recognizing finger motions, generating the corresponding CGH and uploading them to SLM, is ~42 ms. This is equivalent to 23.8 Hz. Since the refresh rate of SLM is 60 Hz which is sufficiently larger than 23.8 Hz, we can realize the 23.8 fps frame rate when uploading CGH on SLM, approaching a real-time and video-level fast modulation of laser beam.

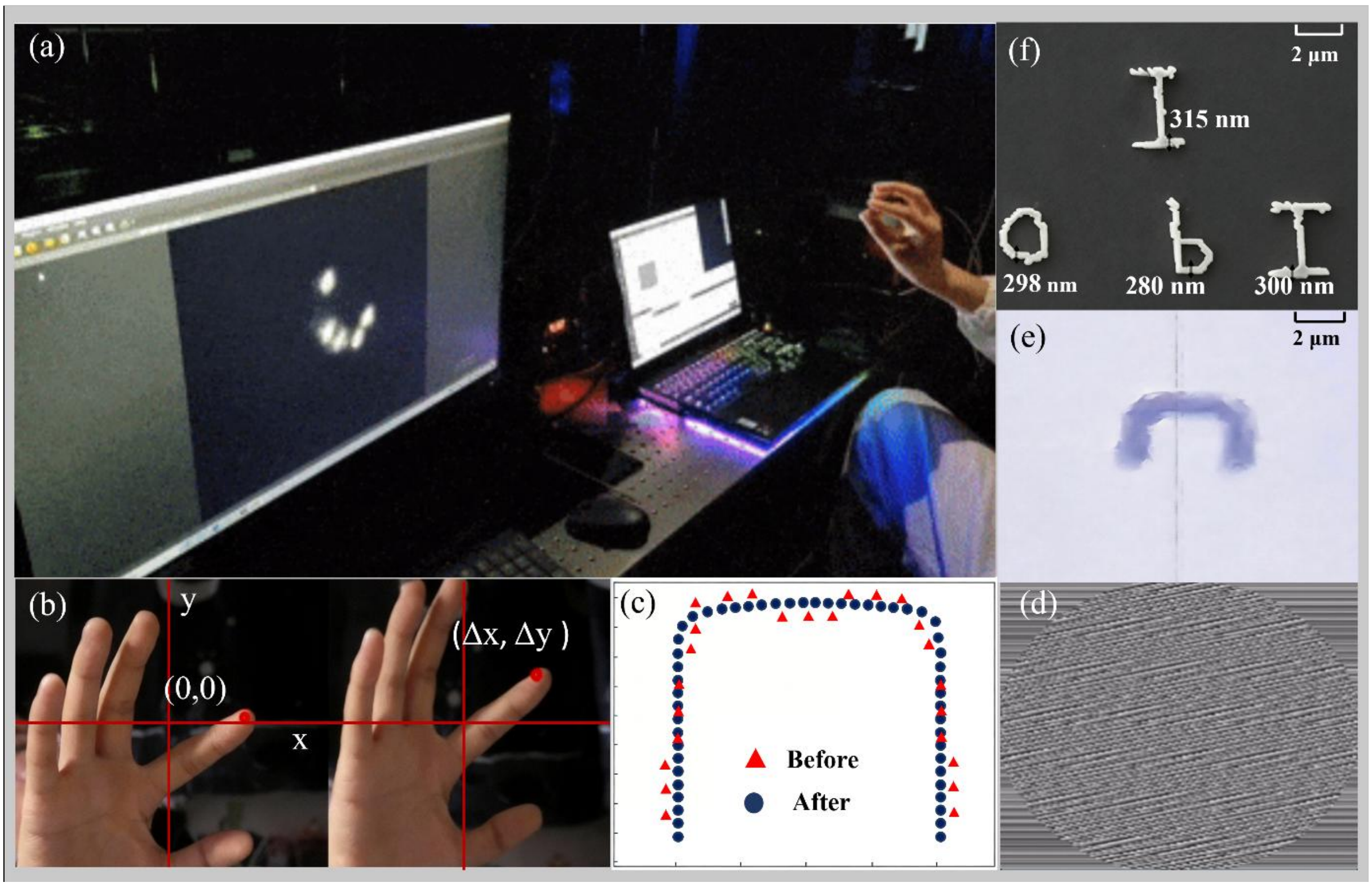

**Fig. 4 Working Diagram of IPALM System, Working Principle and Key Experimental Results.** (a) Live working diagram of the IPALM system.; (b)-(f) Femtosecond laser direct writing based on IPALM technology. (b) The image of the finger movement trajectory on the image detection plane. The red dots represent the results of real-time detection of motion perception; (c) The motion trajectory before and after the four-time moving exponential smoothing median filtering process, where smoothing factor $\alpha = 0.5$ and window size $k = 5$; (d) The real-time holographic phase diagram generated by the SSP algorithm; (e) The actual fabrication result observed using a white field microscope with a 40× objective;(f) SEM image of the fabrication structures at 12 mW power.

## 5.4 Samples and procedures in laser direct writing and cell manipulation

To demonstrate the "mind-to-matter" capability without a programming, we employ the IPALM technique in laser direction writing of nanostructures and optical manipulation of particles.

<u>*Laser direct writing of nanostructures*</u>

In the section, we used a mixture of Pentaerythritol Triacrylate (PETA) monomer and 7-Diethylamino-2-thienoylcoumarin (DETC) photosensitizer as photoresist. DETC is a multipurpose reagent, it absorbs 780 nm femtosecond laser by two-photon absorption (TPA) and emits green fluorescence. Thus, we can monitor the modulated laser beam from the fluorescence directly. PETA and DETC are mixed at a 0.5 wt% ratio. The mixture is further magnetically stirred for 10 hours, then centrifuged to remove bubbles.

A glass slide substrate with low fluorescence is ultrasonically cleaned with acetone-isopropanol-deionized water. The liquid photoresist is uniformly coated on the substrate. After being exposed under the irradiation of femtosecond laser with a power of 5—100 mW depending on the beam number and objective, the photoresist is developed with ethanol for 8 minutes. The final structures can be roughly monitored by white field microscope (Fig. 4e) and finely characterized by scanning electron microscope (SEM, Fig. 4f).

*Optical manipulation of particles*

Laser can direct drive particle motions through different mechanisms. Based on Mie scattering theory, particles in an optical field experience scattering forces ($F_{sca} = Q_{sca} I_0 A/c$)[36] which push the cell along the propagation direction of light, i.e. vertical direction in this investigation. The second one is buoyancy ($F_g = \rho_0 g \beta (T - T_0)$) [37]generated by the photothermal effect influences. It causes vertical motion of cells as well. The third one is gradient force ($F_{grad} = 2\pi\alpha \nabla I_0 / c{n_m}^2$)[38]. In contrast to the former two mechanisms, the gradient force can drive the particles as the laser beam moves. This is the major mechanism we adopt for cell manipulation in horizon planes. The former two mechanisms cannot provide effective control of cell motion in horizontal planes and will induce unexpected vertical motion. Therefore, we restrict all the particles in a microfluidic shallow chamber to restrict the vertical motion.

The microfluidic shallow chamber is fabricated by UV lithography. First of all, the mold of SU-8 photoresist has been generated by UV lithography. Then, the structure is replicated by PDMS which is subsequently bound to a glass cover slide by oxygen plasma. The microfluidic shallow chamber has a rectangular cross section, 50 μm width and 10 μm height. The small height can effectively restrict the vertical motion of particles, providing a venue for transverse cell manipulation.

Two types of particles have been employed to demonstrate the particle manipulation capability of IPALM method in the applications for biomedical engineering and material science. One is sheep blood cell (Zhengzhou Pingrui Biotechnology Co., LTD) (Fig. 3a-g). The samples were centrifuged at 2000 g for 10 minutes to remove plasma, resuspended in PBS, and labeled with DETC (0.1 mg/mL). The other sample is silica microspheres (Fig. 3h-k), which were dispersed in the ethanol-DETC solution by ultrasonication for 10 minutes to achieve a homogeneous suspension for optical trapping studies.

## Supplemental document.

See Supplementary Materials for supporting contents. The supporting video shows the animation of how

IPALM technique works.

## Reference

1 Dou, J. *et al.* Vector modulation of fully-polarized phase conjugate light field through scattering media. *Optics and Laser Technology* **181**, 111987, doi:10.1016/j.Optlastec.2024.111987 (2025).

2 Shao, L., Li, Z., Zhang, Z., Wang, X. & Zhu, W. Multi-Channel Metasurface for Versatile Wavefront and Polarization Manipulation. *Advanced Materials Technologies* **7**, 2200524, doi:https://doi.org/10.1002/admt.202200524 (2022).

3 Zhu, Y. *et al.* Partition laser assembling technique. *Additive Manufacturing* **119**, 105119, doi:https://doi.org/10.1016/j.addma.2026.105119 (2026).

4 Rubinsztein-Dunlop, H. *et al.* Roadmap on structured light. *Journal of Optics* **19**, 013001, doi:10.1088/2040-8978/19/1/013001 (2017).

5 Gori, F., Guattari, G. & Padovani, C. Bessel-Gauss beams. *Optics Communications* **64**, 491-495, doi:https://doi.org/10.1016/0030-4018(87)90276-8 (1987).

6 McGloin, D. & Dholakia, K. Bessel beams: Diffraction in a new light. *Contemporary Physics* **46**, 15-28, doi:10.1080/0010751042000275259 (2005).

7 Waller, L., Situ, G. & Fleischer, J. W. Phase-space measurement and coherence synthesis of optical beams. *Nature Photonics* **6**, 474-479, doi:10.1038/nphoton.2012.144 (2012).

8 Moitra, P. *et al.* Electrically Tunable Reflective Metasurfaces with Continuous and Full-Phase Modulation for High-Efficiency Wavefront Control at Visible Frequencies. *ACS Nano* **17**, 16952-16959, doi:10.1021/acsnano.3c04071 (2023).

9 S., O. A., Carolina, R.-P. & Víctor, A. Generation of the "perfect" optical vortex using a liquid-crystal spatial light modulator. *Optics Letters* **38**, 534-536, doi:10.1364/ol.38.000534 (2013).

10 Li, M., Yan, S., Zhang, Y. & Yao, B. Generation of controllable chiral optical fields by vector beams. *Nanoscale* **12**, 15453-15459, doi:10.1039/D0NR02693J (2020).

11 Ruiter, A. G. T., Moers, M. H. P., van Hulst, N. F. & de Boer, M. Microfabrication of near-field optical probes. *Journal of Vacuum Science & Technology B: Microelectronics and Nanometer Structures Processing, Measurement, and Phenomena* **14**, 597-601, doi:10.1116/1.589142 %J Journal of Vacuum Science & Technology B: Microelectronics and Nanometer Structures Processing, Measurement, and Phenomena (1996).

12 T., O. A. *et al.* Dynamic Cell and Microparticle Control via Optoelectronic Tweezers. *Journal of Microelectromechanical Systems* **16**, 491-499, doi:10.1109/jmems.2007.896717 (2007).

13 Lyu, X. *et al.* Optofluidic three-dimensional microfabrication and nanofabrication. *Nature* **650**, 613-620, doi:10.1038/s41586-025-10033-x (2026).

14 Hoogstrate, A., Drunen, C., Venrooy, B. & Henselmans, R. Manufacturing of high-precision aspherical and freeform optics. *Proceedings of SPIE - The International Society for Optical Engineering* **8450**, 84502Q, doi:10.1117/12.926040 (2012).

15 Tuchin, V. *Tissue Optics: Light Scattering Methods and Instruments for Medical Diagnosis*. (SPIE,

2007).

16 Marin-Palomo, P. *et al.* Microresonator-based solitons for massively parallel coherent optical communications. *Nature* **546**, 274-279, doi:10.1038/nature22387 (2017).

17 Riccardi, M. & Martin, O. J. F. Electromagnetic Forces and Torques: From Dielectrophoresis to Optical Tweezers. *Chemical Reviews* **123**, 1680-1711, doi:10.1021/acs.chemrev.2c00576 (2023).

18 Zhou, C. *et al.* Real-time physical compression computational ghost imaging based on array spatial light field modulation and deep learning. *Optics and Lasers in Engineering* **156**, 107101, doi:https://doi.org/10.1016/j.optlaseng.2022.107101 (2022).

19 Hofmann, J., Scheibinger, R., Fischer, B., Chemnitz, M. & Schmidt, M. A. Programmable liquid-core fibers: Reconfigurable local dispersion control for computationally optimized ultrafast supercontinuum generation. *Nature Communications* **16**, 7918, doi:10.1038/s41467-025-63213-8 (2025).

20 Wang, C. *et al.* Real-time, label-free phase contrast imaging via programmable SLM-based phase mask in off-axis digital holography. *Opt. Express* **33**, 51402-51416, doi:10.1364/OE.573704 (2025).

21 Kong, D., Cao, L., Shen, X., Zhang, H. & Jin, G. Image Encryption Based on Interleaved Computer-Generated Holograms. *IEEE Transactions on Industrial Informatics* **14**, 673-678, doi:10.1109/TII.2017.2714261 (2018).

22 Porfirev, A. Spatial-light-modulator-assisted laser manipulation in air. *Optical Engineering* **59**, 055109, doi:10.1117/1.Oe.59.5.055109 (2020).

23 Han, W. *et al.* 3D Femtosecond Laser Beam Deflection for High-Precision Fabrication and Modulation of Individual Voxelated PCM Meta-Atoms. *Advanced Science* **12**, 2413316, doi:https://doi.org/10.1002/advs.202413316 (2025).

24 Li, Z., Allegre, O. & Li, L. Realising high aspect ratio 10 nm feature size in laser materials processing in air at 800 nm wavelength in the far-field by creating a high purity longitudinal light field at focus. *Light: Science & Applications* **11**, 339, doi:10.1038/s41377-022-00962-x (2022).

25 Tringides, C. M. & Mooney, D. J. Materials for Implantable Surface Electrode Arrays: Current Status and Future Directions. *Advanced Materials* **34**, 2107207, doi:https://doi.org/10.1002/adma.202107207 (2022).

26 Constantinescu, C., Deepak, K. L. N., Delaporte, P., Utéza, O. & Grojo, D. Arrays of metallic micro-/nano-structures by means of colloidal lithography and laser dewetting. *Applied Surface Science* **374**, 124-131, doi:10.1016/j.apsusc.2015.10.073 (2016).

27 Xu, H. *et al.* Real-time capable feedrate optimization for laser processes with redundant axes via two-stage regularized linear programming. *International Journal of Machine Tools and Manufacture* **213**, 104342, doi:https://doi.org/10.1016/j.ijmachtools.2025.104342 (2025).

28 Christopher, J. B., Michael, G. O. & Gorby, A. D. Validation of an analytical solution for depth of correlation in microscopic particle image velocimetry. *Measurement Science and Technology* **15**, 318, doi:10.1088/0957-0233/15/2/002 (2004).

29 Hu, C.-L., Wang, L., Chen, M.-L. & Pei, C. A real-time interactive decision-making and control framework for complex cyber-physical-human systems. *Annual Reviews in Control* **57**, 100938, doi:https://doi.org/10.1016/j.arcontrol.2024.100938 (2024).

30 Jiang, Z. *et al.* De-coherent parallel laser processing of ultradense nanopores for high-density, large-area

3D optical phase encoding. *Nature Communications* **17**, 1070, doi:10.1038/s41467-025-67828-9 (2026).

31 Wang, M. *et al.* Fusing Stretchable Sensing Technology with Machine Learning for Human–Machine Interfaces. *Advanced Functional Materials* **31**, 2008807, doi:https://doi.org/10.1002/adfm.202008807 (2021).

32 Geng, Y. Q., Lai, F. L., Luo, H. & Gao, F. Nmix: a hybrid deep learning model for precise prediction of 2'-O-methylation sites based on multi-feature fusion and ensemble learning. *Briefings in Bioinformatics* **25**, bbae601, doi:10.1093/bib/bbae601 (2024).

33 P., S., G., S. K. & K.S., V. B. Music Recommendation System Using Facial Emotions. *Advances in Science and Technology* **124**, 44-52, doi:10.4028/p-4s4w34 (2023).

34 Ditzinger, T. & Haken, H. The impact of fluctuations on the recognition of ambiguous patterns. *Biological Cybernetics* **63**, 453-456, doi:10.1007/BF00199577 (1990).

35 Yueqiang, Z., Wei, Z., Chen, Z., Kaige, W. & Jintao, B. Non-iterative multifold strip segmentation phase method for six-dimensional optical field modulation. *Optics Letters* **47**, 1335-1338, doi:10.1364/ol.444419 (2022).

36 He, L. *et al.* Solid Particle Swarm Measurement in Jet Fuel Based on Mie Scattering Theory and Extinction Method. *Sensors* **23**, 2837 (2023).

37 Pan, D. *et al.* Transparent Light-Driven Hydrogel Actuator Based on Photothermal Marangoni Effect and Buoyancy Flow for Three-Dimensional Motion. *Advanced Functional Materials* **31**, 2009386, doi:https://doi.org/10.1002/adfm.202009386 (2021).

38 Zheng, H., Chen, H., Ng, J. & Lin, Z. Optical gradient force in the absence of light intensity gradient. *Physical Review B* **103**, 035103, doi:10.1103/PhysRevB.103.035103 (2021).

## Acknowledgement

We appreciate the valuable suggestions from Prof. Chao-Kuei Lee. This investigation is supported by National Natural Science Foundation of China (Grant No. 51927804, 61775181, 61378083).